\documentclass[11pt,twoside]{article}
\usepackage{asp2010}

\resetcounters

\begin{document}

\title{Diagnosing small- and large-scale structure in the winds of
  hot, massive OB-stars}
\author{J.O.~Sundqvist,$^1$ \& S.P.~Owocki,$^1$
\affil{$^1$Dept. of Physics \& Astronomy, University of Delaware, USA}}

\begin{abstract}

It is observationally as well as theoretically well established that
the winds of hot, massive OB-stars are highly structured on a broad
range of spatial scales. This paper first discusses consequences of
the \textit{small-scale} structures associated with the strong
instability inherent to the line-driving of these winds. We
demonstrate the importance of a proper treatment of such wind clumping
to obtain reliable estimates of mass-loss rates, and also show that
instability simulations that are perturbed at the lower boundary
indeed display significant clumping quite close to the wind base, in
general agreement with observations.

But a growing subset of massive stars has also been found to possess
strong surface magnetic fields, which may channel the star's outflow
and induce also \textit{large-scale} wind structures and cyclic
behavior of spectral diagnostics. The paper concludes by showing that
multi-dimensional, magneto-hydrodynamical wind simulations, together
with detailed radiative-transfer modeling, can reproduce remarkably
well the periodic Balmer line emission observed in slowly rotating
magnetic O stars like HD\,191612.

\end{abstract}

\section{Introduction}

Hot, massive OB-stars have strong winds, with typical mass-loss rates
$\dot{M} \approx 10^{-6} \, \rm M_\odot /yr$ and terminal speeds
$v_\infty \approx 2\,000 \, \rm km/s$, driven by metal-line scattering
of the star's intense continuum radiation field. The first
quantitative description of such line-driving was given in the seminal
paper by \citep[][CAK]{Castor75}, who assumed a steady-state,
spherically symmetric, and homogeneous outflow. And since then,
extensions and applications of this CAK theory have had considerable
success in explaining many of the observed gross properties and global
trends of hot star winds, such as the relation between the wind
momentum and the stellar luminosity and the basic metallicity
dependence of mass-loss rates. Nevertheless, it has over the years
also become very clear that, actually, these winds are neither
steady-state, nor homogeneous; rather they seem to be highly
structured on a broad range of spatial scales. (For a very
comprehensive review of hot star winds, we refer the reader to
\citealt{Puls08}.)

This paper summarizes our ongoing efforts in characterizing the nature
and consequences of such wind structure. After first briefly reviewing
the background to the strong instability inherent to line driving, we
next concentrate on confronting theoretical predictions of the
resulting \textit{small-scale} structure with observations, and
examine what consequences such wind clumping have for the interpretation
of spectral diagnostics. The paper then continues by discussing how
the presence of a strong surface magnetic field may channel the
outflow along closed field lines, and how the \textit{large-scale}
wind structures predicted by simulations of such magnetically confined
winds may be diagnosed using the rotational phase variation of optical
recombination lines like H$\alpha$.

\section{Small-scale structure and wind clumping} 

The amount of evidence supporting a clumped stellar wind is nowadays
overwhelming. \citet{Sundqvist11b} give a comprehensive overview of
the current status; this paper concentrates on two items, namely i)
the theoretical background and ii) the effects of clumping on wind
diagnostics used to derive mass-loss rates and clumping properties.

\subsection{Theory: \ \ The line-deshadowing instability} 
\label{LDI}

\begin{figure}
\begin{minipage}{6.8cm}
\resizebox{\hsize}{!}
{\includegraphics[angle=90]{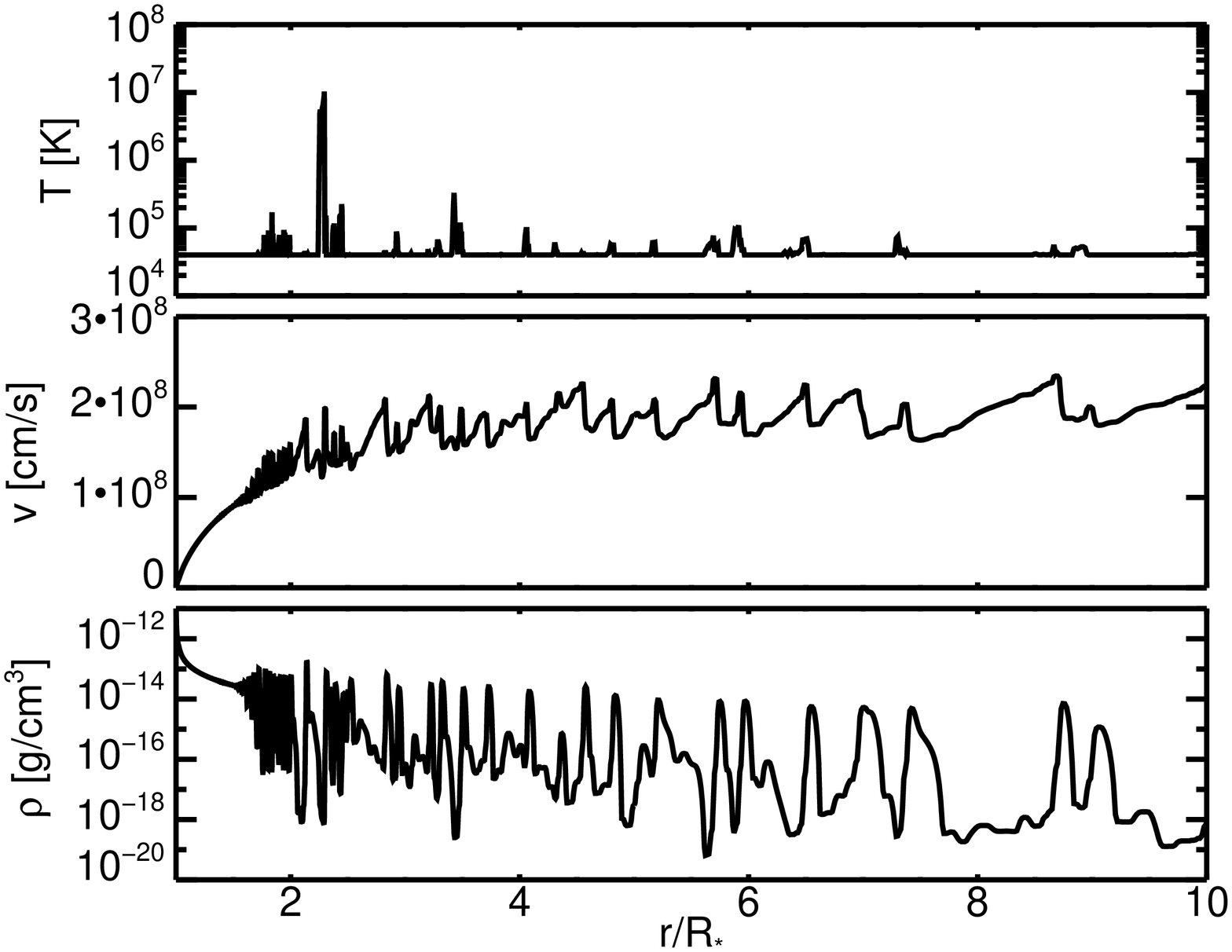}}
\end{minipage}
\begin{minipage}{6.8cm}
\resizebox{\hsize}{!}
{\includegraphics[angle=90]{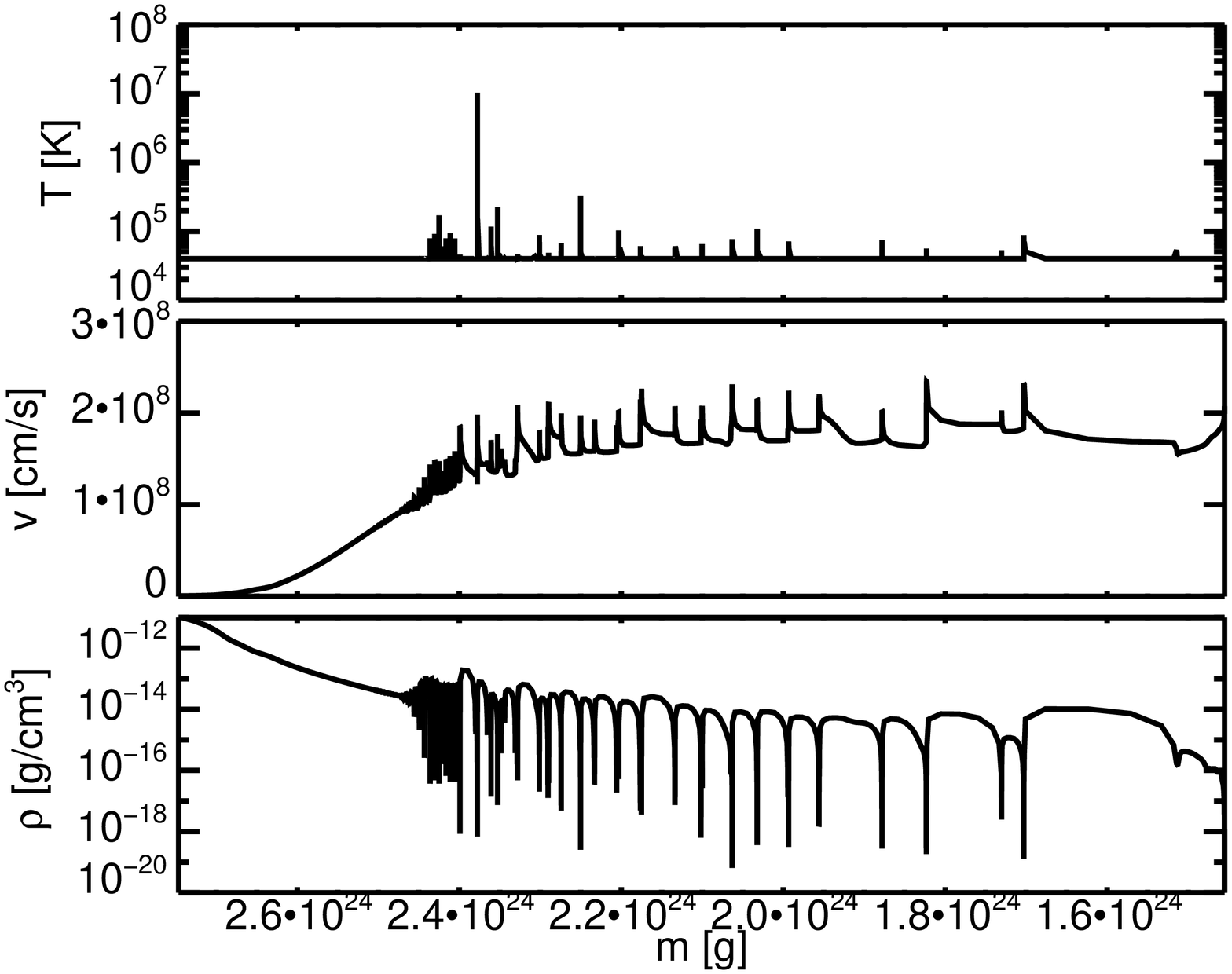}}
\end{minipage}
\caption{Snapshot of temperature, velocity, and density in a 1-D LDI
  simulation, plotted versus radius (left) and a Lagrangian mass
  coordinate (right), as defined in e.g. \citet{Owocki88} and
  illustrating how little wind mass is contained in the high-speed
  rarefactions.}
\label{Fig:snaps}
\end{figure}

Linear stability analyses showed already during the 80s
\citep{Owocki84, Owocki85} that the line force responsible for
accelerating hot star winds is highly unstable on spatial scales near
and below the Sobolev length $L = v_{\rm th}/(dv/dr) \approx (v_{\rm
  th}/v) r \approx 0.01 r$, where $v_{\rm th} \approx 10 \, \rm km/s$
is the ion thermal speed and $v \approx 1000 \, \rm km/s$ is a typical
wind flow speed. Direct numerical modeling \citep{Owocki88,
  Feldmeier95, Dessart05} has since confirmed that the non-linear
growth of this \textit{line-deshadowing instability} (LDI) leads to
high speed rarefactions that steepen into strong reverse shocks,
whereby most wind material is compressed into spatially very narrow
``clumps'' (or shells in 1-D simulations). Fig.~\ref{Fig:snaps}
illustrates this characteristic structure, which forms the basis for
our current understanding and interpretation of wind clumping.
Further, the presence of strong shocks and hot gas in LDI simulations
also provide a generally accepted explanation for the soft X-rays
observed from OB-stars by orbiting telescopes like {\sc chandra}
\citep[][see also Cohen, this volume]{Feldmeier97}.

The simulation snapshot displayed in Fig.~\ref{Fig:snaps} has been
calculated using the numerical hydrodynamics code VH-1 (developed by
J.~Blondin et al.), with the radiation line force computed following
the so-called smooth source function \citep[SSF,][]{Owocki96}
approach. The SSF method allows one to follow the non-linear evolution
of the LDI, while also accounting for the stabilizing line-drag effect
\citep{Lucy84} of the diffuse, scattered radiation field. In the
displayed simulation, this line drag exactly cancels the LDI at the
stellar surface, leading to marginal stabilization of the wind
base. However, as mass parcels move away from the stellar surface, the
relative influence of the line-drag decreases and structure starts to
develop, roughly at $r \approx 1.5 R_\star$.

\begin{figure}
\begin{minipage}{6.8cm}
\resizebox{\hsize}{!}
{\includegraphics[angle=90]{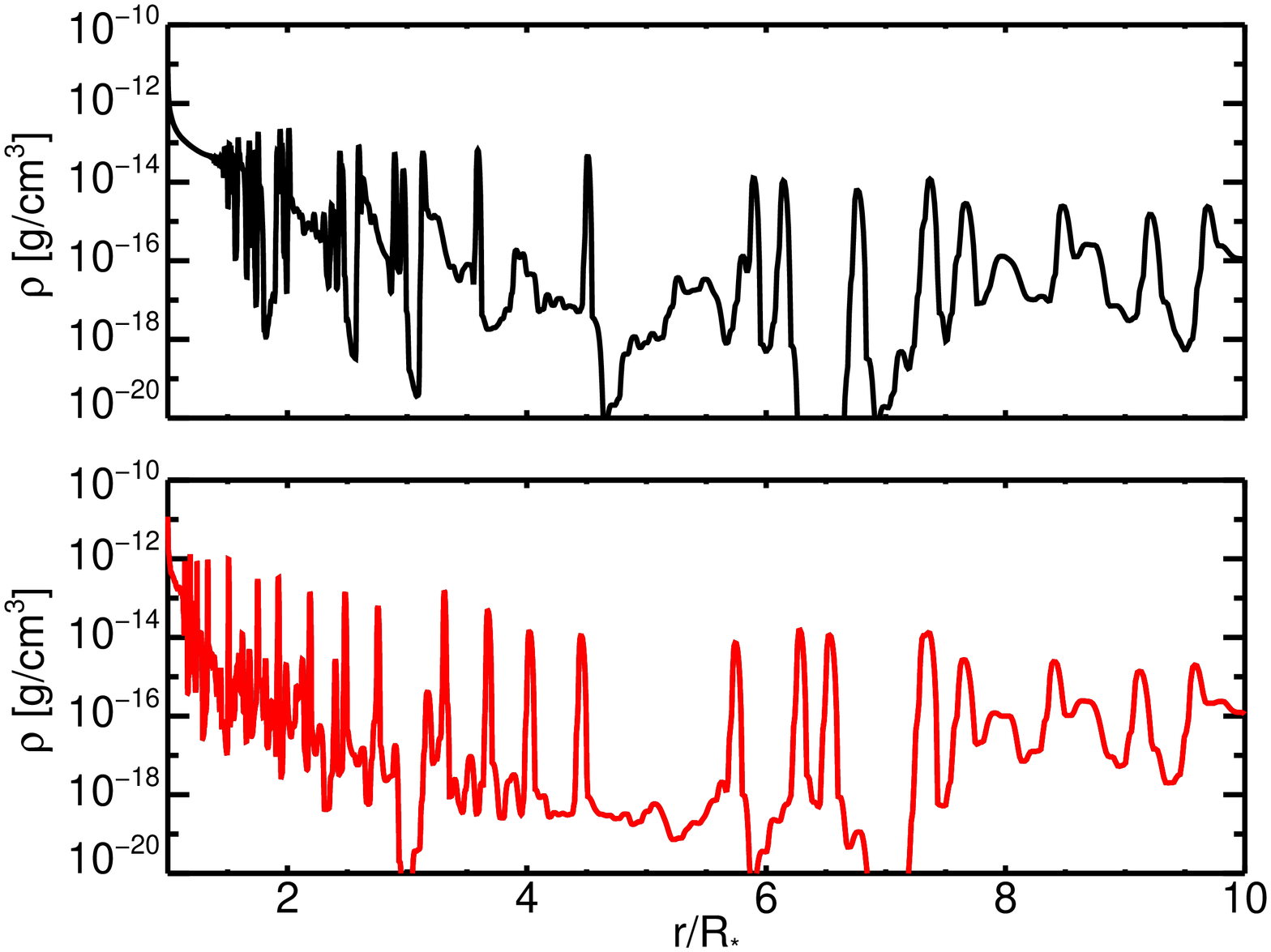}}
\end{minipage}
\begin{minipage}{6.8cm}
\resizebox{\hsize}{!}
{\includegraphics[angle=90]{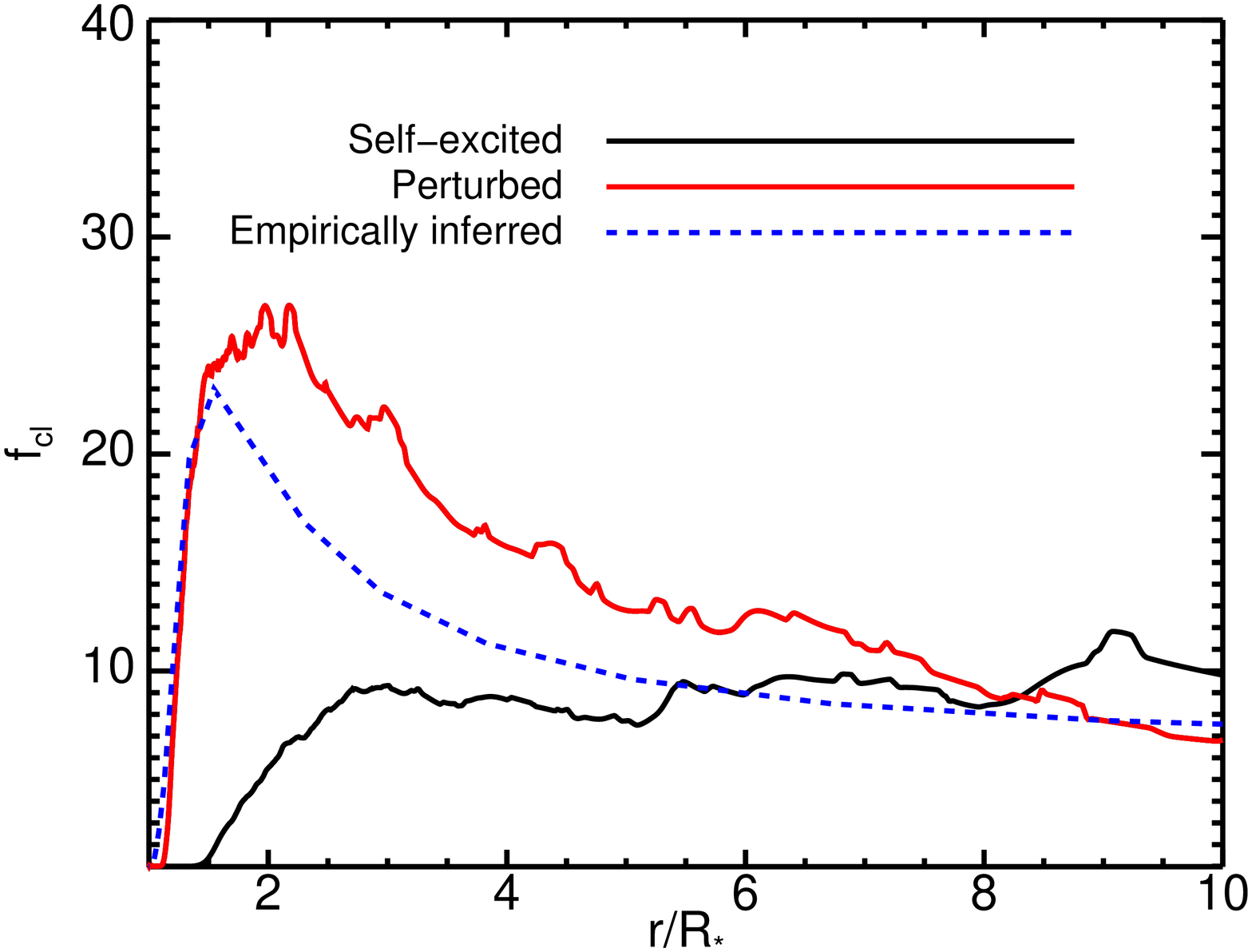}}
\end{minipage}
\caption{\textbf{Left:} Density snapshots for an LDI simulation with
  self-excited structure (upper) and one in which the lower boundary
  has been perturbed by a sound wave (lower). \textbf{Right:} Radial
  stratification of the clumping factor as derived for $\zeta$ Pup by
  \citet{Najarro11} vs.  theoretical predictions.}
\label{Fig:fcl}
\end{figure}

But note that the exact amount of such instability-damping in the
lower wind depends critically on delicate details of the competition
between the LDI (arising from direct absorption of stellar continuum
radiation) and the line-drag (arising from the scattering source
function), and so is quite sensitive to, e.g., external
perturbations. Indeed, we report here on additional model computations
in which we perturb the density $\rho_0$ at the lower boundary by
introducing a generic sound wave of amplitude $\delta \rho/\rho_0 =
0.25$ and period 4000 sec. And such simulations now display
significant structure much closer to the wind base, as shown by
Fig.~\ref{Fig:fcl}. The right panel of this figure compares the radial
stratification of the time-averaged clumping factor $f_{\rm cl} \equiv
\langle \rho^2 \rangle/\langle \rho \rangle^2$ in simulations with
self-excited structure and a perturbed lower boundary, respectively,
and shows clearly how the latter model is more strongly clumped in the
inner wind. This result is in general good agreement with
observations, as further discussed at the end of
Sect.~\ref{diagnose1}.

\subsection{Diagnostics: \ \  
  Deriving mass-loss rates and clumping properties of OB-star winds} 
\label{diagnose1}

\begin{figure}
\begin{minipage}{6.8cm}
\resizebox{\hsize}{!}
{\includegraphics[angle=90]{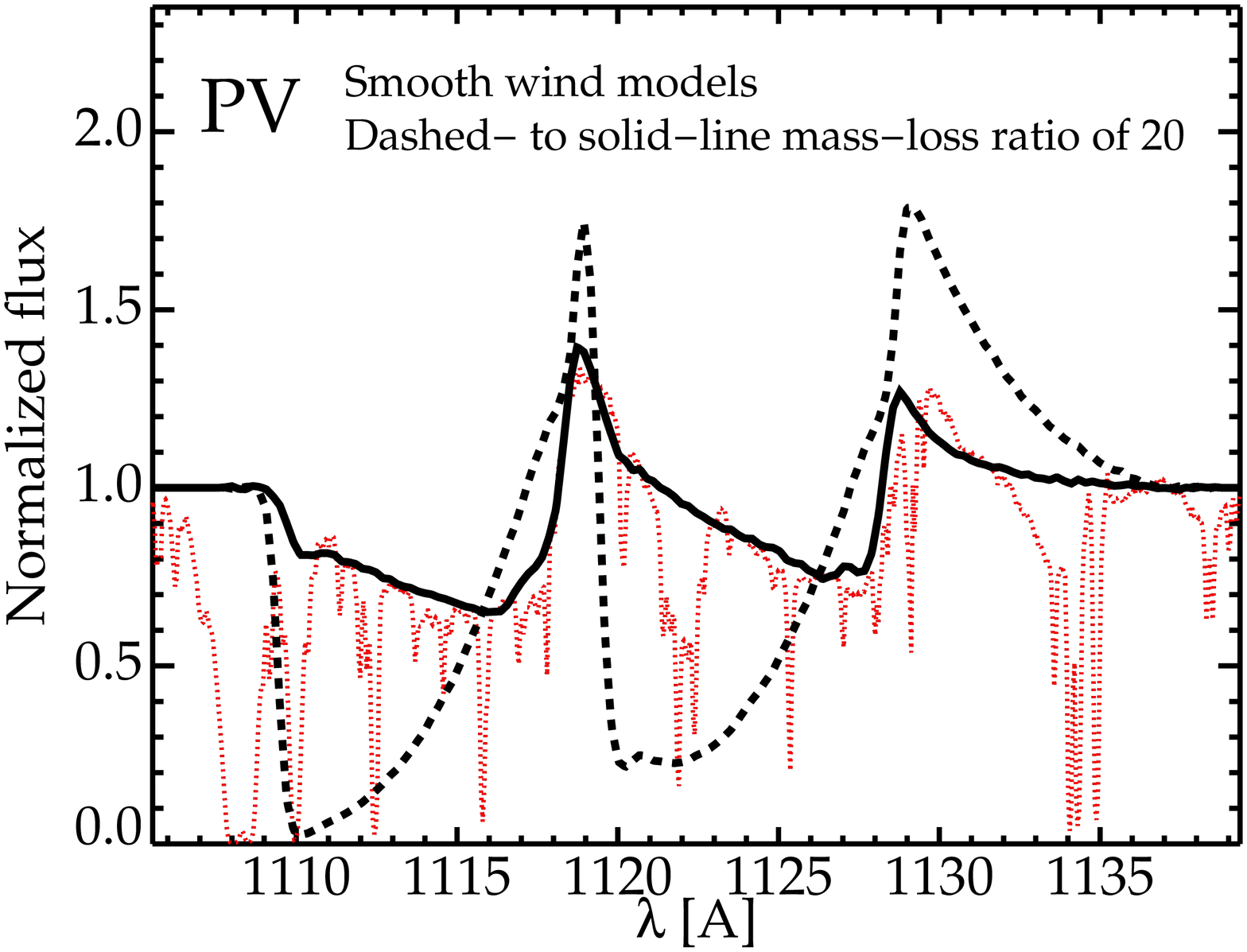}}
\end{minipage}
\begin{minipage}{6.8cm}
\resizebox{\hsize}{!}
{\includegraphics[angle=90]{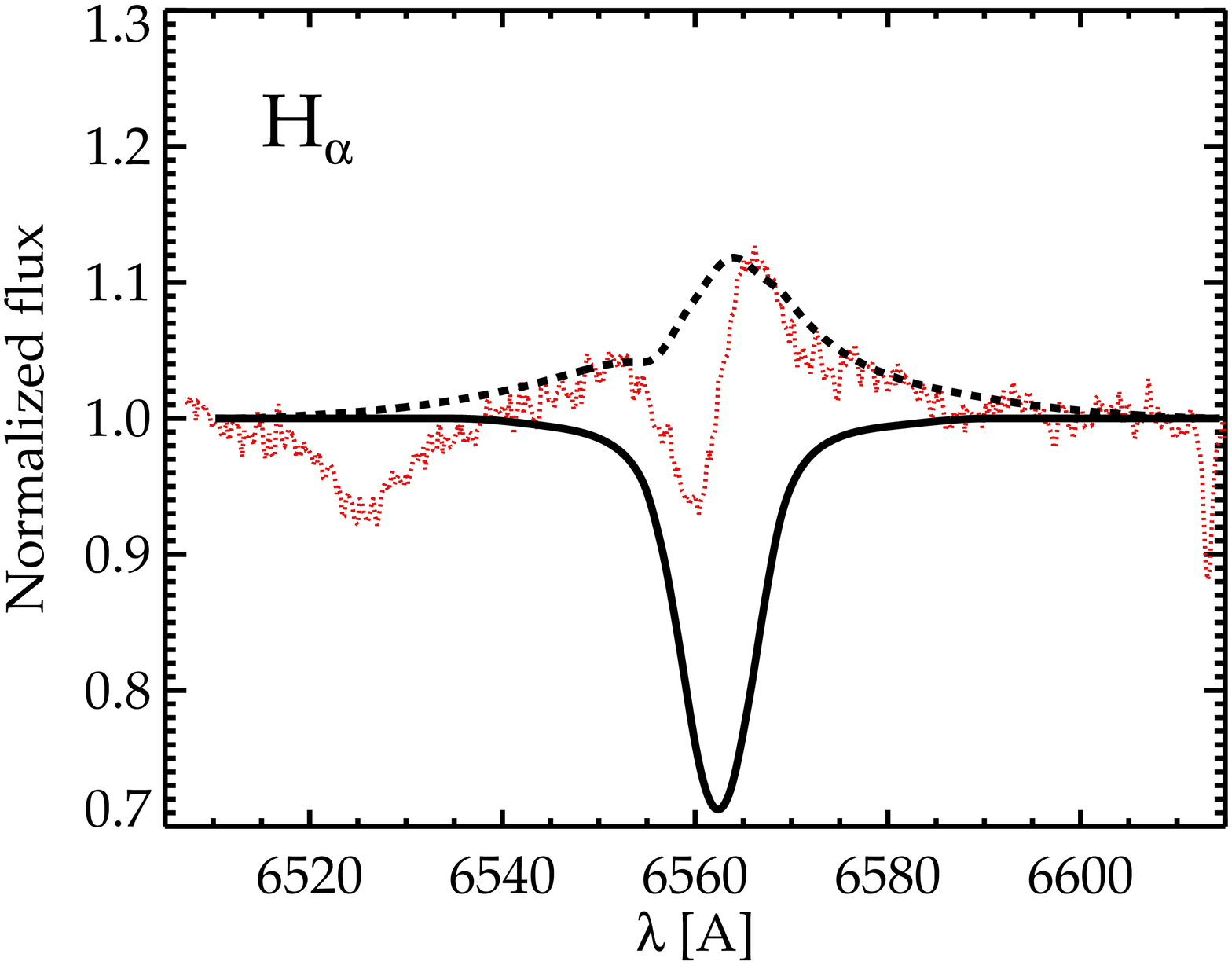}}
\end{minipage}
\caption{Observed (dotted red lines) and synthetic (black solid and
  dashed lines) PV and H$\alpha$ line profiles in $\lambda$ Cep.
  Observations from \citet{Fullerton06} and
  \citet{Markova05}. Synthetic profiles calculated using the unified
  (photosphere+wind) NLTE model atmosphere code {\sc fastwind}
  \citep{Puls05} and the Monte-Carlo code developed by
  \citet{Sundqvist10, Sundqvist11}. All profiles were calculated
  assuming a \textit{smooth} wind, and illustrate how the mass-loss
  rate ($6 \times 10^{-6} \, \rm M_\odot /yr$) required to fit
  H$\alpha$ is $\sim 20$ times higher than that required to fit the PV
  lines.}
  \label{Fig:pv_ha_smooth}
\end{figure}

The average mass-loss rate from a time-dependent LDI simulation
actually is quite similar to that of its steady-state CAK start model,
suggesting the resulting wind structure only has secondary
effects\footnote{from, e.g., a changed wind ionization balance
  \citep{Muijres11}.} on theoretical mass-loss predictions. On the
  other hand, the structure in density and velocity seriously affects
  the radiative transfer through the wind, and thereby also the
  interpretation of \textit{observed} spectra.

To illustrate this, let us compare mass-loss rates derived from
different wind diagnostics using \textit{smooth} wind models. Taking
the Galactic O6 supergiant $\lambda$ Cep as an example,
Fig.~\ref{Fig:pv_ha_smooth} shows how the rate derived from the
optical recombination line H$\alpha$ is $\sim 20$ times higher than
the rate derived from the UV resonance lines of phosphorus {\sc v}
(PV), the latter also being almost an order of magnitude lower than
predicted by the line-driven wind theory! Such inconsistencies have
been found by many studies \citep[with perhaps the most prominent
  example being the study of PV in 40 Galactic O stars
  by][]{Fullerton06}, and are likely a consequence of neglecting wind
clumping when deriving the mass-loss rates.

Most diagnostic studies of clumping have assumed that the wind
consists of statistically distributed, \textit{optically thin} clumps
embedded in a void background medium, and neglected any disturbances
on the velocity field.  The most important result of such optically
thin clumping is that the opacities of diagnostics depending on the
square of the density, such as H$\alpha$ and thermal infra-red/radio
emission, are higher than in a smooth wind model with the same
mass-loss rate. The scaling invariant is $\sqrt{f_{\rm cl}} \dot{M}$,
leading to overestimates of mass-loss rates derived from smooth wind
models by $\sqrt{f_{\rm cl}}$. In contrast, diagnostics depending only
linearly on density, such as UV resonance lines and bound-free
attenuation of X-rays, are not directly affected by optically thin
clumping\footnote{they can be indirectly affected though, through a
  modified ionization balance \citep[e.g.][]{Bouret03}}.  Combined
optical/infra-red/radio studies \citep{Puls06, Najarro11} using this
approach typically show that mass-loss rates derived from H$\alpha$
and smooth wind models should be scaled down by factors $2 \dots
5$. Such reductions have also been confirmed by independent studies of
bound-free X-ray attenuation \citep[][see also Cohen, this
  volume]{Cohen10, Cohen11}, and would imply modest reductions of
present-day theoretical predictions of mass-loss rates, as well as
clumping factors on order $f_{\rm cl} \approx 10$. 

\paragraph{Clumping in the inner wind} While such characteristic 
clumping factors are in general accordance with theoretical
predictions, diagnostic studies \citep{Eversberg98, Bouret03, Puls06,
  Cohen11} also suggest the wind is significantly structured well
below the $r \approx 1.5 R_\star$ predicted for the onset of clumping
by standard LDI simulations with self-excited structure. The right
panel of Fig.~\ref{Fig:fcl} compares the radial stratification of
$f_{\rm cl}$ derived for $\zeta$ Pup by the comprehensive
multi-diagnostic study by \citet{Najarro11} with theoretical
predictions. The figure shows that perturbing the lower boundary (see
Sect.~\ref{LDI}) can indeed induce significant wind structure also in
the lower wind, leading then to quite good overall agreement with
observations. A journal paper discussing such base perturbations, as
well as other effects that may affect the predicted structure 
in the inner wind, is currently in preparation.

\begin{figure}
\begin{minipage}{6.8cm}
\resizebox{\hsize}{!}
{\includegraphics[angle=90]{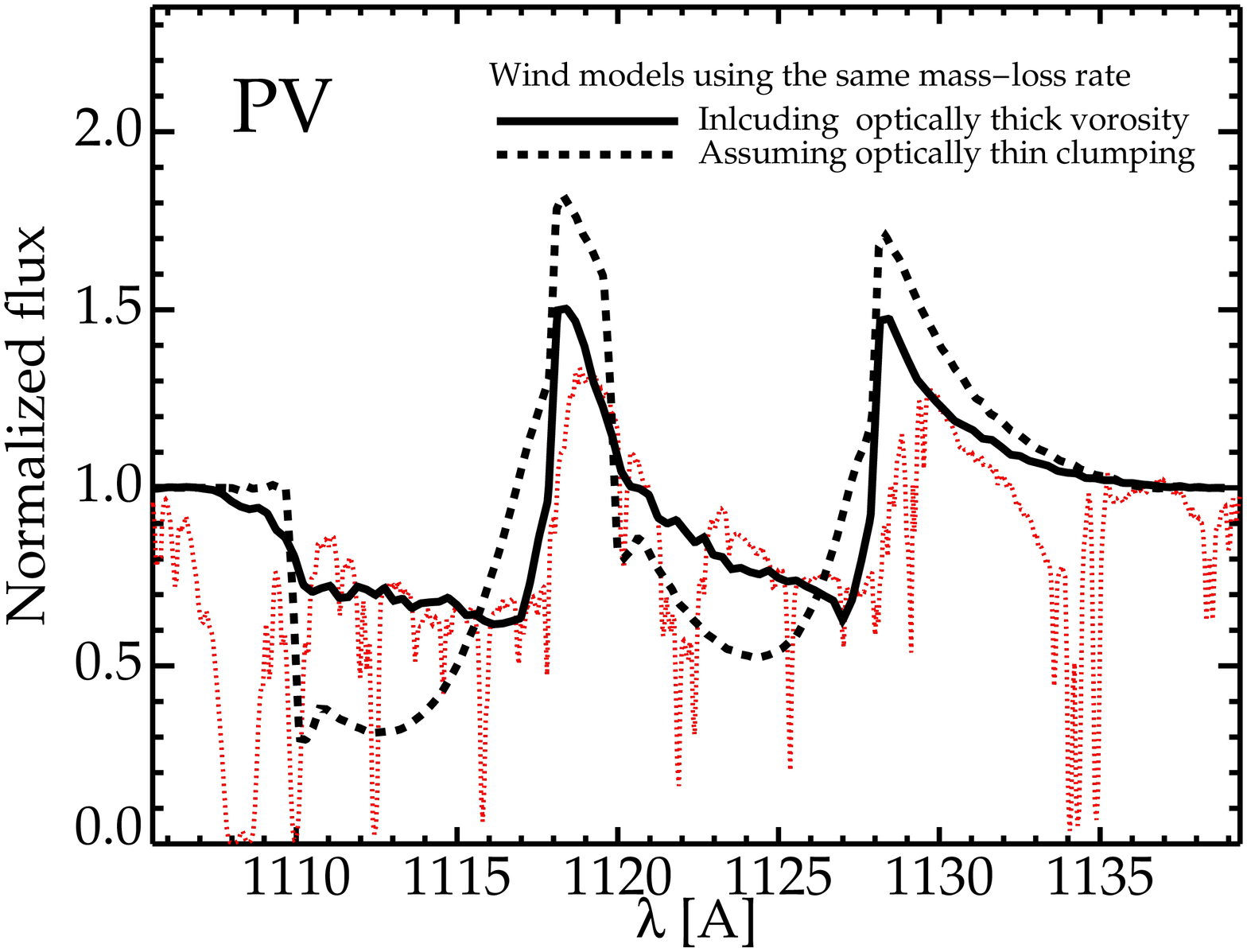}}
\end{minipage}
\begin{minipage}{6.8cm}
\resizebox{\hsize}{!}
{\includegraphics[angle=90]{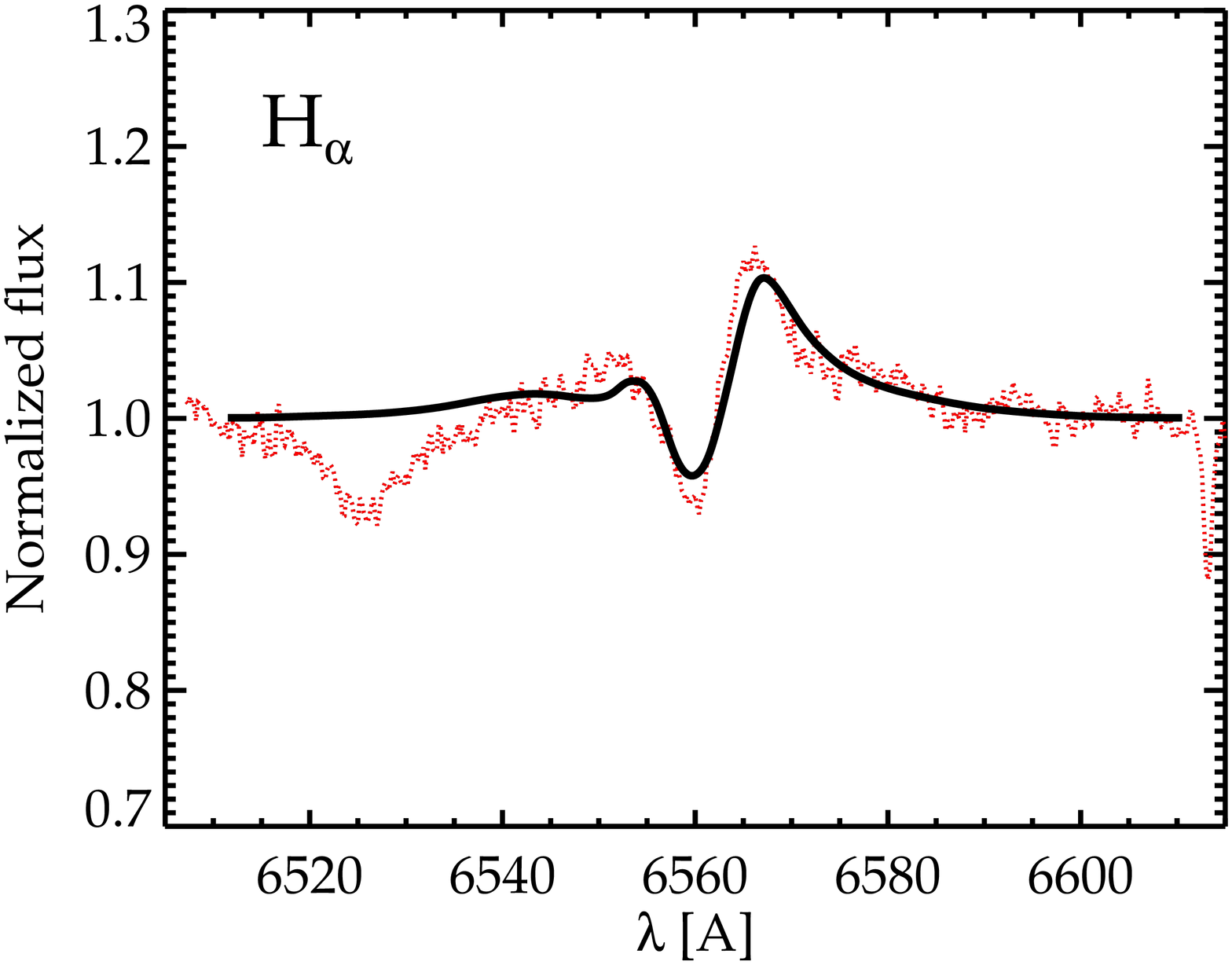}}
\end{minipage}
\caption{Same as Fig.~\ref{Fig:pv_ha_smooth}, but now accounting for
  clumping and optically thick vorosity (solid lines, see text), and
  using a single mass-loss rate $1.5 \times 10^{-6} \, \rm
  M_\odot/yr$. As an illustration of the importance of optically thick
  clumping for UV resonance lines, the dashed lines show synthetic PV
  profiles calculated with the same mass-loss rate but now under the
  assumption of \textit{optically thin} clumps (i.e. neglecting
  vorosity).}
\label{Fig:pv_ha}
\end{figure}

\paragraph{Optically thick clumping -- porosity and vorosity} 
But note that such optically thin clumping does not really help
explain the unexpected weakness of the PV lines displayed in
Fig.~\ref{Fig:pv_ha}, as these line are basically unaffected by this
approach. It is important to realize, however, that if the assertion
of optically thin clumps is not met for the investigated process,
additional effects become important in the radiative transfer. For
continuum diagnostics, self-shielding of opacity within optically
thick clumps leads to increased escape of photons through porous
channels in between the clumps.  Such \textit{porosity} thus has the
general effect of reducing the wind's effective opacity
\citep{Feldmeier03, Owocki04}. Though it has been suggested porosity
might be important for X-ray line attenuation in hot star winds
\citep{Feldmeier03, Oskinova06}, recent studies show that this is
quite unlikely as there is no apparent evidence for significant
porosity in the data \citep[][Leutenegger et al., submitted]{Cohen08}
  and it would further imply mean free paths between clumps much
  longer than predicted by theory \citep{Owocki06, Sundqvist12}.

But the situation is quite different for the inherently very strong UV
resonance \textit{lines}, in which clumps easily can become optically
thick \citep{Oskinova07}. Due to the narrow Doppler width of the line
profile, line photons in a rapidly expanding stellar wind can only
interact with the plasma over a very narrow spatial range. Therefore
many such line photons will simply leak through the wind via
``porous'' channels in the velocity field, without ever interacting
with the optically thick clumps. Hence this effect has been dubbed
velocity porosity, or ``vorosity'' \citep{Owocki08}. And in analogy
with spatial porosity, the main effect of such vorosity is to reduce
the wind's effective opacity.

Using models accounting for both clumping and optically thick
vorosity, Fig.~\ref{Fig:pv_ha} again shows PV and H$\alpha$ fits to
$\lambda$ Cep. And in contrast to before
(Fig.~\ref{Fig:pv_ha_smooth}), both diagnostics can now be well fit
using the same mass-loss rate, which now is only a modest factor of
two lower than the theoretical prediction by \citet{Vink00}. This
illustrates the importance of a proper treatment of wind clumping when
deriving mass-loss rates from OB-stars.

\section{Large-scale wind structure due to magnetic fields}      

Let us now switch focus from the small-scale structures associated
with the LDI to the more large-scale wind structures associated with
the presence of a strong surface magnetic field. Due to recent
advances in spectropolarimetric techniques, such large-scale fields
are being detected in OB-stars at an increasing rate (for an overview
of the current observational status, see e.g. Wade, this volume).

\subsection{Theory: \ \  Dynamical magnetospheres} 

\begin{figure}
\begin{minipage}{4.5cm}
\resizebox{\hsize}{!}
{\includegraphics[]{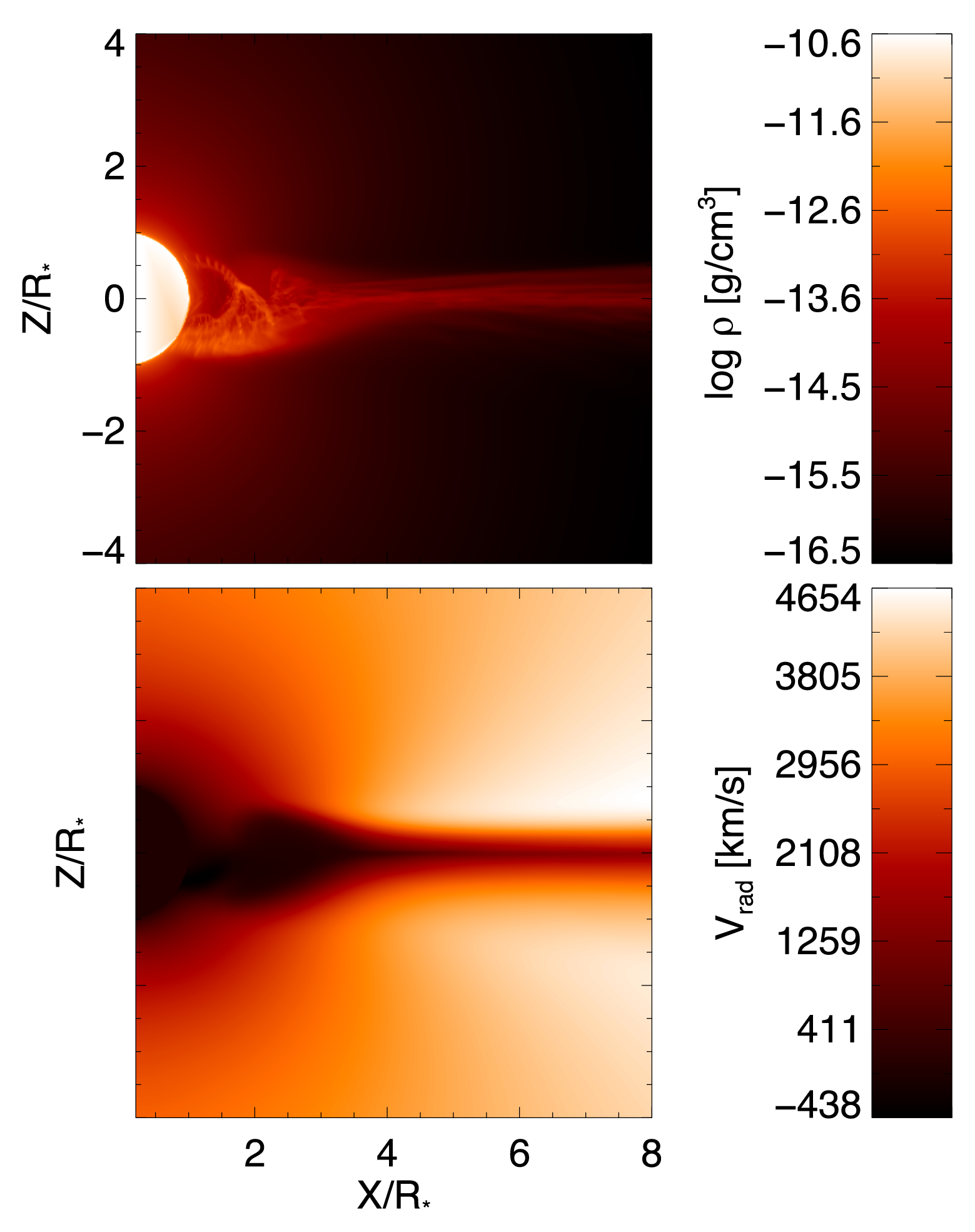}}
\end{minipage}
\begin{minipage}{8.5cm}
\resizebox{\hsize}{!}
{\includegraphics[angle=90]{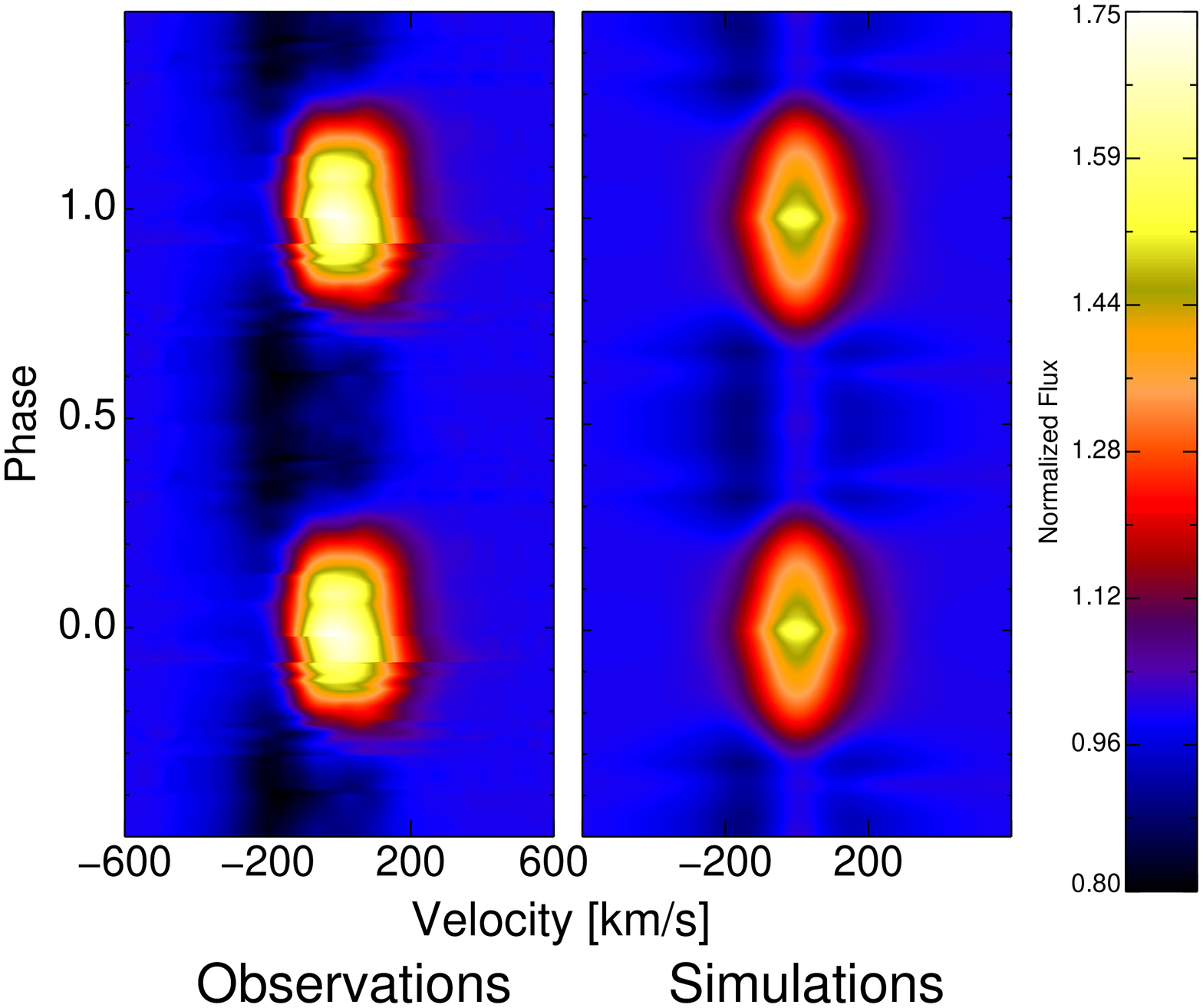}}
\end{minipage}
\caption{\textbf{Left:} Contour maps of density (upper) and radial
  velocity (lower) in an MHD wind simulation, obtained by averaging over
  $\sim 100$ snapshots well after initiation. \textbf{Right:} Observed
  \citep[ephemeris from][]{Howarth07} and synthetic H$\alpha$ dynamic
  spectra for HD19162, calculated as described in the text and in
  \citet{Sundqvist12b}.}
\label{Fig:ha_dm}
\end{figure}

The effectiveness of the magnetic field in channeling the stellar wind
outflow may be characterized by the ratio of the magnetic to wind
kinetic energy density:
\begin{equation}
	\eta = \frac{B^2/(8 \pi)}{\rho v^2/2} = \eta_\star   
	\frac{(r/R_\star)^{-4}}{v/v_\infty},  
	\label{Eq:eta} 
\end{equation}
where the second equality assumes a dipolar field and defines the
so-called wind confinement parameter $\eta_\star = (B_\star^2
R_\star^2)/(\dot{M} v_\infty)$ \citep{udDoula02}, with $B_\star$ the
equatorial surface field strength. A key point is that for $\eta_\star
> 1$, the Alfv\'{e}n radius $R_{\rm A} \approx \eta_\star^{1/4}
R_\star$ at which the magnetic and wind kinetic energy densities are
equal is located away from the stellar surface, allowing then for wind
plasma to be channeled along closed field loops toward the magnetic
equator. But note also that the steeper radial decline of the magnetic
energy density than the wind kinetic energy density inevitably means
the wind will always dominate at large enough radii (as is directly
evident from the second equality in eqn.~\ref{Eq:eta}) and force the
field lines to open up and essentially follow the radial flow.

Magneto-hydrodynamical (MHD) simulations following the time-evolution
of the competition between the magnetic field and the radiatively
driven wind confirms this basic picture \citep{udDoula02}. The MHD
simulation (computed by A. ud-Doula) used to synthesize the H$\alpha$
emission in the following section is described in \citet{Sundqvist12b}
and assumes $\eta_\star = 50$. The left panels of Fig.~\ref{Fig:ha_dm}
show that below the Alfv\'{e}n radius $R_{\rm A} \approx 2.7 R_\star$
the wind indeed becomes trapped and the material is pulled back onto
the star by gravity. But despite this very dynamic behavior, with gas
constantly in upward and downward motion, the transient suspension of
plasma within such closed field loops leads to a region around the
magnetic equator that statistically is overdense.

Note that this structure is physically distinct from that predicted
for rapidly rotating magnetic stars characterized by $R_{\rm A} >
R_{\rm K}$ \citep{Townsend05, udDoula08}, where $R_{\rm K}$ is the
Kepler corotation radius at which the centrifugal force balances
gravity. Fur such stars, centrifugal forces can support any trapped
material above $R_{\rm K}$, which allows for accumulation of wind
plasma over very long time-scales and to the formation of a
\textit{centrifugal magnetosphere} (CM). In contrast, the
characteristic structure described above, appropriate for slowly
rotating magnetic stars with $R_\star < R_{\rm A} < R_{\rm K}$, forms
a \textit{dynamical magnetosphere} (DM). An important distinction
between a CM and a DM is that the latter accumulates wind plasma only
over the relatively short dynamical time-scale, implying that such DMs
require the quite high mass-loss feeding rate of an O star in order to
emit in H$\alpha$ \citep[][see also Petit \& Owocki, in
  prep.]{Sundqvist12b}.

\subsection{Diagnostics: \ \ 
  Rotational phase variation of Balmer line emission} 
 
We compute synthetic H$\alpha$ profiles directly from the MHD
simulation discussed above by solving the formal integral of the
radiative transfer equation in a 3-D cylindrical coordinate system
aligned toward the observer. Since we target slowly rotating stars, we
may neglect rotational effects on the wind dynamics and use the same
simulation for any obliquity angle $\beta$ between the rotation and
magnetic axes. \citet{Sundqvist12b} give a full description of the
computation method. The right panel of Fig.~\ref{Fig:ha_dm} compares
the rotational phase variation of such synthetic H$\alpha$ spectra
with observations of the very slowly rotating \citep[$P = 538\, \rm
  days$, $v_{\rm rot} \approx 1 \, \rm km/s$,][]{Howarth07} magnetic O
star HD\,191612, using an inclination angle $i = 50 \deg$ and $\beta =
50 \deg$, consistent with the magnetic geometry constraints derived by
\citet{Wade11a}. Fig.~\ref{Fig:ha_dm} shows that both the peak flux at
phase 0 and the extended minimum around phase 0.5 are well
reproduced. The time modulation is caused by the observer's changing
projected surface area of overdense H$\alpha$ emitting wind plasma as
the star rotates. At maximum emission (phase 0), the DM is viewed
essentially from above the magnetic pole, whereas at minimum (phase
0.5) it is viewed almost edge-on. Thus the DM in the latter phase has
a smaller projected surface area, leading to weaker emission.

The good overall agreement between observations and simulations
obtained for HD\,191612 is encouraging, and provides strong support
for the basic concept of the dynamical magnetosphere model. We are
currently in the process of extending the modeling presented here to
more magnetic O stars with $R_\star < R_{\rm A} < R_{\rm K}$ that also
show periodic Balmer emission, like HD\,57682 (Grunhut et al.,
submitted to MNRAS) and $\theta^1$ Ori C.

\acknowledgements J.O.S. gratefully acknowledges funding from NASA ATP
grant NNX11AC40G

\bibliography{talk_sundqvist}

\end{document}